# Leveraging Raman response in X-cut thin-film lithium tantalate for ultrabroadband combs and polychromatic visible light

**Xin Wang[1,†], Mingkun Xiao[1,†], Min Sun[1,†], Ronghong Gao[2], Yuqi Chen[1], Zhengshun Lei[1], Xun Zhang[1], Wenfeng Zhou[1], Jintian Lin[2], Yikai Su[1,*], Xingchen Ji[1,*], Yong Zhang[1,*]**

1. State Key Lab of Photonics and Communications, Department of Electronic Engineering, Shanghai Jiao Tong University, Shanghai 200240, China
2. The Extreme Optoelectromechanics Laboratory (XXL), School of Physics, East China Normal University, Shanghai 200241, China

*yikaisu@sjtu.edu.cn,xingchenji@sjtu.edu.cn,yongzhang@sjtu.edu.cn;

†These authors contributed equally.

## Abstract

X-cut thin-film lithium tantalate (TFLT) offers a unique combination of $\chi^{(3)}$ nonlinearity, electro-optic effects, and a high optical damage threshold. However, its strong Raman response has historically hindered broadband Kerr comb generation. Here, we leverage this inherent Raman response by engineering coupling-defined dissipation. This allows us to reconfigure the relative thresholds of Raman and Kerr processes without modifying the intrinsic microring dispersion. Through this coupling-engineered threshold control, we can deliberately access distinct comb states, ranging from pure Kerr combs to Raman–Kerr synergistic broadband combs. We demonstrate a Kerr comb spanning 450 nm and a Raman–Kerr comb spanning 650 nm, representing the broadest combs reported to date on X-cut TFLT platforms. Moreover, in strongly coupled devices, we show that a single near-infrared pump can generate visible emission across multiple bands (from violet to red) via cascaded $\chi^{(2)}$ sum-frequency processes. Our work demonstrates that a strong Raman response can be transformed from a parasitic competitor into an enabling mechanism for achieving broader comb spectra and generating polychromatic light. This work establishes X-cut TFLT as a powerful monolithic platform for nonlinear light sources, electro-optic functions, and complex photonic systems.

## Introduction

Microresonator frequency combs are driving integrated photonics from single-wavelength sources toward controllable, multifunctional frequency comb systems[1,2], supporting applications such as optical computing[3], sensing[4], communications[5], radar[6,7], optical clocks[8], and microwave photonics[7]. However, many of these applications require strong ($\chi^{(2)}$) second-order nonlinearity for key functions like high-speed electro-optic modulation and frequency conversion—a property that is absent in most mature microcomb platforms[9-11]. Consequently, these platforms cannot monolithically combine $\chi^{(2)}$ and $\chi^{(3)}$ functionalities, forcing either heterogeneous material integration[12-15] or multi-chip assemblies[16-18], both of which introduce coupling loss, fabrication complexity, manufacturing cost, and system-level instability, hindering further on-chip integration[11,19]. Therefore, a monolithic platform that concurrently enables Kerr comb generation (via $\chi^{(3)}$) and $\chi^{(2)}$-based functions is urgently needed.

Thin-film lithium niobate (TFLN) and thin-film lithium tantalate (TFLT) are promising candidates for the monolithic $\chi^{(2)}+\chi^{(3)}$ platform sought in comb-driven integrated photonics, as they combine a broad transparency window, low propagation loss, high electro-optic coefficients, and third-order nonlinearities (Fig. 1a, b)[2], together with recent advances in wafer-scale thin-film processing. To date, the platforms have enabled electro-optic modulators with bandwidths exceeding 250 GHz[20] and high-performance Kerr optical frequency combs[21,22]. However, modulators are typically implemented on the X-cut because of its larger electro-optic coefficient $r_{33}$, whereas most high-performance Kerr combs have been demonstrated on the Z-cut owing to its weaker Raman response and more favorable dispersion properties. As a result, current comb-driven photonic systems require two different crystal cuts, preventing monolithic integration[16]. Achieving monolithic integration, therefore, demands the realization of high-quality frequency combs directly on X-cut platforms.

A major obstacle on X-cut TFLN and TFLT is their strong Raman response, which allows stimulated Raman scattering to compete with Kerr four-wave mixing and thereby alters comb formation[23,24]. Stable Kerr-comb operation has therefore typically required Raman suppression, pursued via two principal strategies. The first, demonstrated on Z-cut TFLN, uses dissipation engineering to raise the Raman threshold—by introducing extra loss in the Raman gain region[25] or shifting resonances away from the Raman peak through free-spectral-range design[21]. Yet X-cut platforms exhibit a substantially stronger Raman response[26], and such loss-engineering approaches remain unexplored there. The second strategy employs racetrack resonators aligned along specific crystal orientations to minimize Raman gain via crystallographic anisotropy[23,26-28]. While effective for Raman suppression, this orientation constraint narrows the design space for broadband dispersion engineering. Separately, bend-induced avoided mode crossings have been used in X-cut racetrack resonators to assist Kerr-comb initiation and improve coherence and stability, but they do not directly extend the comb bandwidth. Consequently, simultaneously managing Raman suppression, dispersion engineering and comb bandwidth on X-cut platforms remains a significant challenge. Notably, compared with TFLN, TFLT offers a higher optical damage threshold, lower birefringence and weaker DC drift. These advantages promise broader comb generation, thus providing a viable pathway towards monolithically integrated comb-driven photonics on the X-cut platform.

Here, we demonstrate that the Raman response in X-cut TFLT can be utilized rather than suppressed,

by combining dispersion engineering with coupling-defined dissipation engineering. With the main microring geometry unchanged and thus the intrinsic dispersion preserved, coupling engineering enables different devices to selectively operate in Kerr-dominated, Raman-dominated, and broadband Raman–Kerr comb regimes (Fig. 1c). We achieve stable ultrabroadband Kerr comb generation spanning 450 nm, and further realize Raman-assisted Kerr combs by maximizing the Raman gain while exploiting the high optical damage threshold of TFLT. This yields simultaneous Stokes and anti-Stokes Raman scattering and produces a Raman–Kerr comb spanning 650 nm. To our knowledge, these represent the broadest Kerr and Raman–Kerr combs reported on X-cut TFLN/TFLT platforms. Beyond infrared comb generation, second-order nonlinear processes give rise to multi-band visible emission, including yellow and orange bands, offering a new route to on-chip multi-wavelength visible light generation. These results show that the strong Raman response of X-cut TFLT can be harnessed rather than avoided, establishing a foundation for the monolithic integration of nonlinear comb sources, multi-channel visible emission, and complex photonic functions on the X-cut platform.

## Results

### Dissipation engineering

Modifying the resonator waveguide directly changes both dispersion and loss, making it difficult to tune the thresholds independently. To decouple these effects, we keep the resonator dispersion fixed and engineer only the coupling region. Specifically, we introduce a wavelength-dependent external loss by varying the bus–ring gap $g$, coupling angle $\theta_{\text{couple}}$, and waveguide widths $w_{\text{bus}}$ and $w_{\text{ring}}$. Curved coupling is particularly effective because its extended interaction length and continuously varying phase-matching condition give rise to a strongly wavelength-dependent external coupling rate $\kappa_{ex}= \kappa_{ex}(\lambda, g, w_{\text{bus}}, w_{ring}, \theta_{\text{couple}})$[25]. As a result, the pump mode, Raman Stokes modes, and Kerr sidebands experience different external losses, which shifts their relative thresholds and promotes distinct nonlinear pathways. We begin with the coupled-mode equation for the pump amplitude. The steady-state intracavity pump intensity satisfies: $|a_p|^2 \propto \kappa_{ex,p}/\kappa_p^2 \cdot P_{\text{in}}$ where $P_{\text{in}}$ is the input pump power, $\kappa_p= \kappa_{i,p}+\kappa_{ex,p}$ is the total loss rate of the pump mode, with $\kappa_{i,p}$ the intrinsic loss rate and $\kappa_{ex,p}= \kappa_{ex}(\lambda, g, w_{\text{bus}}, w_{ring}, \theta_{\text{couple}})$ the coupling-controlled external loss rate. The thresholds for Raman and Kerr processes can then be derived as (see Supplementary Note1 for the detailed derivation)

$$P_{\text{th,R}} \propto \frac{\kappa_s \kappa_p^2}{\kappa_{ex,p} G_R}, \quad P_{\text{th,K}} \propto \frac{\kappa_\mu \kappa_p^2}{\kappa_{ex,p} G_K \Phi_\mu}. \tag{1.1}$$

Here $\kappa_s= \kappa_{i,s} + \kappa_{ex,s}$ is the total loss rate of the Raman Stokes mode, $\kappa_\mu$ is the total loss rate of the Kerr sideband mode, $G_R$ and $G_k$ are the effective Raman and Kerr gain coefficients, respectively, and $\Phi_\mu$ is the phase-matching penalty factor set by the detuning and integrated dispersion.

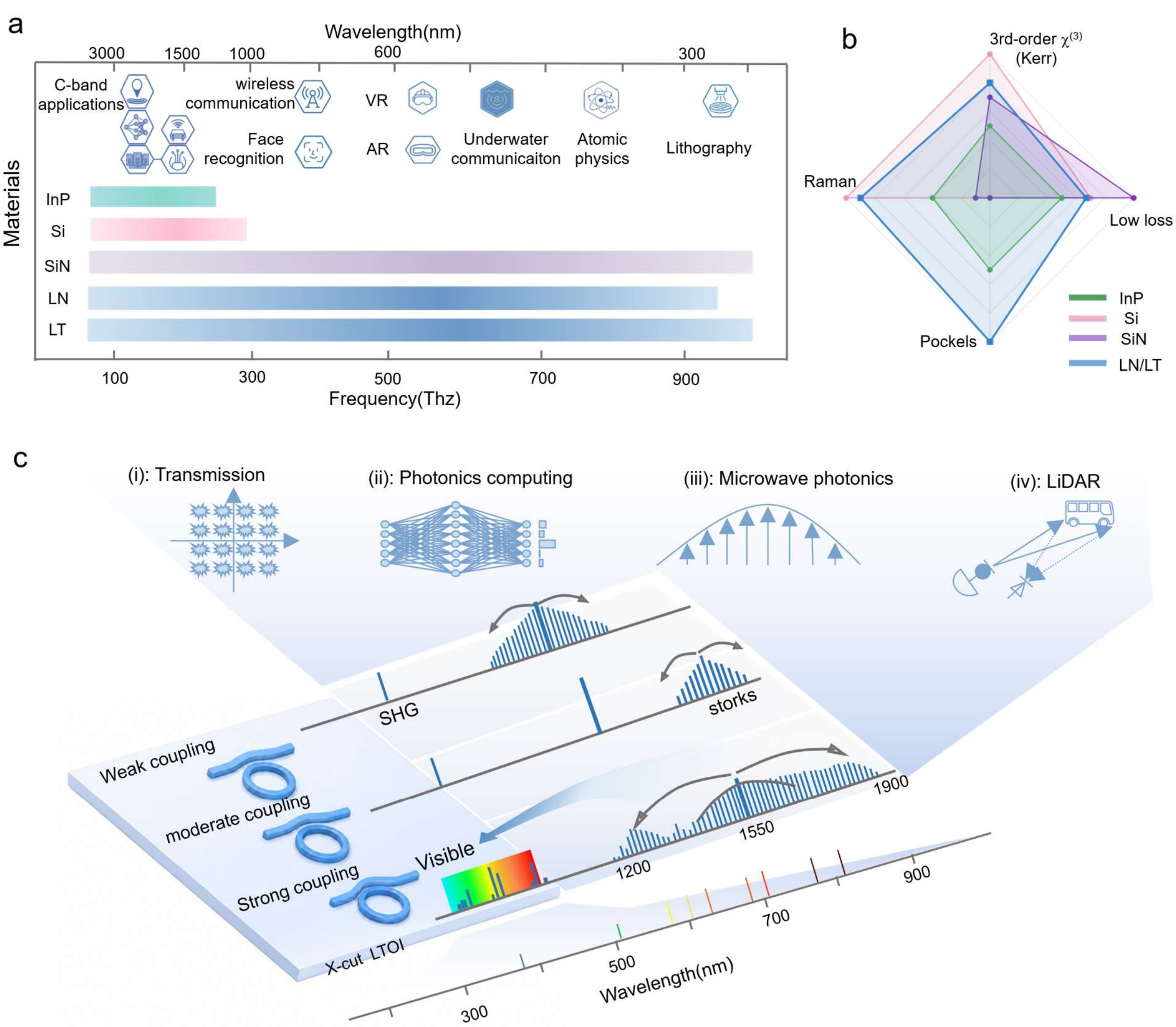

**Fig. 1 | Typical applications of integrated photonic chips and coupling-controlled nonlinear states. a,** Spectral coverage of fully integrated photonic chips and representative applications across different optical bands. Colored bars indicate the transparency windows of five representative material platforms (i.e., InP, Si, SiN, TFLN and TFLT). **b,** Radar chart comparing key performance metrics of TFLN, TFLT, Si, InP, and SiN, including propagation loss, Raman gain, electro-optic coefficient, and third-order nonlinearities. **c,** Coupling-controlled access to distinct nonlinear regimes under identical cavity dispersion. By varying only the coupling condition, three representative comb states are obtained: a Kerr comb, a Raman comb and a Raman-assisted Kerr comb accompanied by multi-band visible emission. Representative applications of optical frequency combs, including optical communications, optical computing and microwave photonic radar, are illustrated in the upper insets.

The coupling condition thus governs the threshold hierarchy via pump injection efficiency, modal loss rates and phase matching. Different coupling configurations can reorder the Raman and Kerr thresholds: a lower Kerr threshold favours a Kerr-comb-dominated state, whereas reducing the loss of a specific Raman Stokes mode produces a discrete Raman or hybrid state. Under strong wavelength-dependent coupling, the Raman branch reaches threshold, seeds localized comb lines, and cooperates with Kerr-induced comb filling to form a broadband Raman–Kerr synergistic frequency comb.

To validate this mechanism, we fixed the microring geometry and varied only the coupling region, calculating the external coupling rate under different conditions (Fig. 2a: 30° coupling angle, 1.3 μm bus width; Fig. 2b: 45°, 1.25 μm)[29]. Near 1550 nm, the coupling in Fig. 2b is stronger. Around 1700 nm (Raman shift 599 $cm^{-1}$), the two parameter sets differ markedly; the second set exhibits a much stronger variation in coupling Q, indicating strong modulation of the loss condition for this Raman channel.

Based on the generalized Lugiato–Lefever equation (LLE), the Raman response is incorporated into the nonlinear processes via a convolution term. By feeding both the coupling rate and the simulated dispersion into the equation, we performed frequency comb simulations. Figures 2c and 2d correspond to the parameter sets of Fig. 2a and Fig. 2b, respectively. For the parameters corresponding to the weak-coupling regime in Fig. 2c, the entire spectrum remains a Kerr comb. Figure 2d demonstrates that, under different gap distances, the spectrum transitions from a Kerr comb to a Raman comb or a Raman–Kerr comb. The schematic inset compares the relative thresholds of the Kerr and Raman processes: when the Kerr threshold is lower than the Raman threshold, the output is dominated by a Kerr comb; as the Kerr threshold rises above the Raman threshold, the system can transition into Raman lasing or a Raman–Kerr comb. These results indicate that the coupling design can selectively rearrange the loss conditions for different Raman channels, thereby altering the threshold competition between Raman oscillation and Kerr four-wave mixing.

## Device characterization

A dissipation-engineered microring resonator of TFLT achieves intrinsic quality factors on the order of $10^6$ and anomalous dispersion, enabling Kerr comb generation. The experimental setup used for device characterization is shown in Fig. 3a. The dissipation-engineered microring resonators are fabricated on the 600-nm-thick X-cut thin-film lithium tantalate-on-insulator (LTOI) wafer. The coupling region of the device is shown in Fig. 3b, and a scanning electron microscope (SEM) image of the waveguide cross-section is presented in Fig. 3c, revealing a sidewall angle of approximately 70°. The microring geometry is carefully designed to support anomalous group velocity dispersion, a key condition for Kerr comb generation. The resonator has a radius of 100 μm, an etch depth of 450 nm, and a top waveguide width of 2.1 μm. We characterize the passive performance of a representative device; fitting a typical resonance peak yields an intrinsic quality factor $Q_{int}=2.06\times10^6$ (Fig. 3d). Figure 3e summarizes the statistical distribution of intrinsic quality factors for two typical classes of devices–weakly coupled and strongly coupled–showing that the intrinsic Q remains predominantly on the order of $10^6$ for both types. This indicates that varying the coupling condition mainly modulates the loading loss of the cavity rather than its intrinsic optical quality.

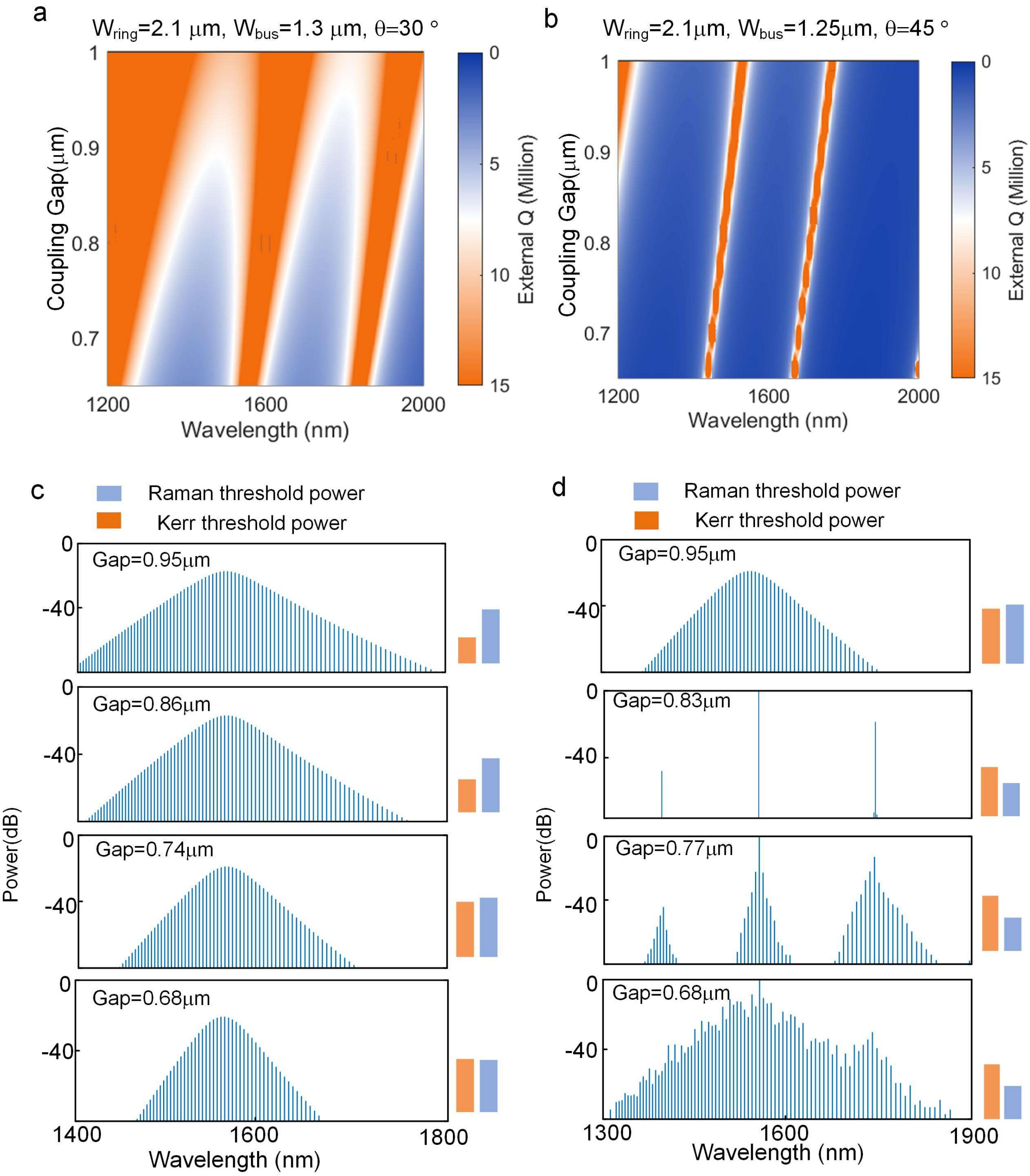

**Fig. 2 | Calculated frequency comb generation states under different coupling rates. a,** External coupling Q as a function of gap and wavelength for a coupling angle of 30° and a bus waveguide width of 1.3 μm. **b,** External coupling Q as a function of gap and wavelength for a coupling angle of 45° and a bus waveguide width of 1.25 μm. **c,** Simulated spectra corresponding to the four coupling parameters indicated in a; insets compare the Raman and Kerr thresholds.

**d,** Simulated spectra corresponding to the four coupling parameters indicated in b, the schematic insets provide a schematic comparison of the relative thresholds. The simulations demonstrate that varying the thresholds can switch the system among a Raman laser, a Kerr comb, and a Raman-Kerr comb.

We further extract the device dispersion from the measured resonance frequencies. As shown in Fig. 3f, the measured integrated dispersion $D_{\mathrm{int}}$ agrees well with the fitted curve, where the reference mode $\mu$=0 corresponds to approximately 1550 nm. The fitted second-order dispersion is $D_2/2\pi$ = -0.354 MHz, confirming that the device operates in the anomalous dispersion regime and supports Kerr comb generation.

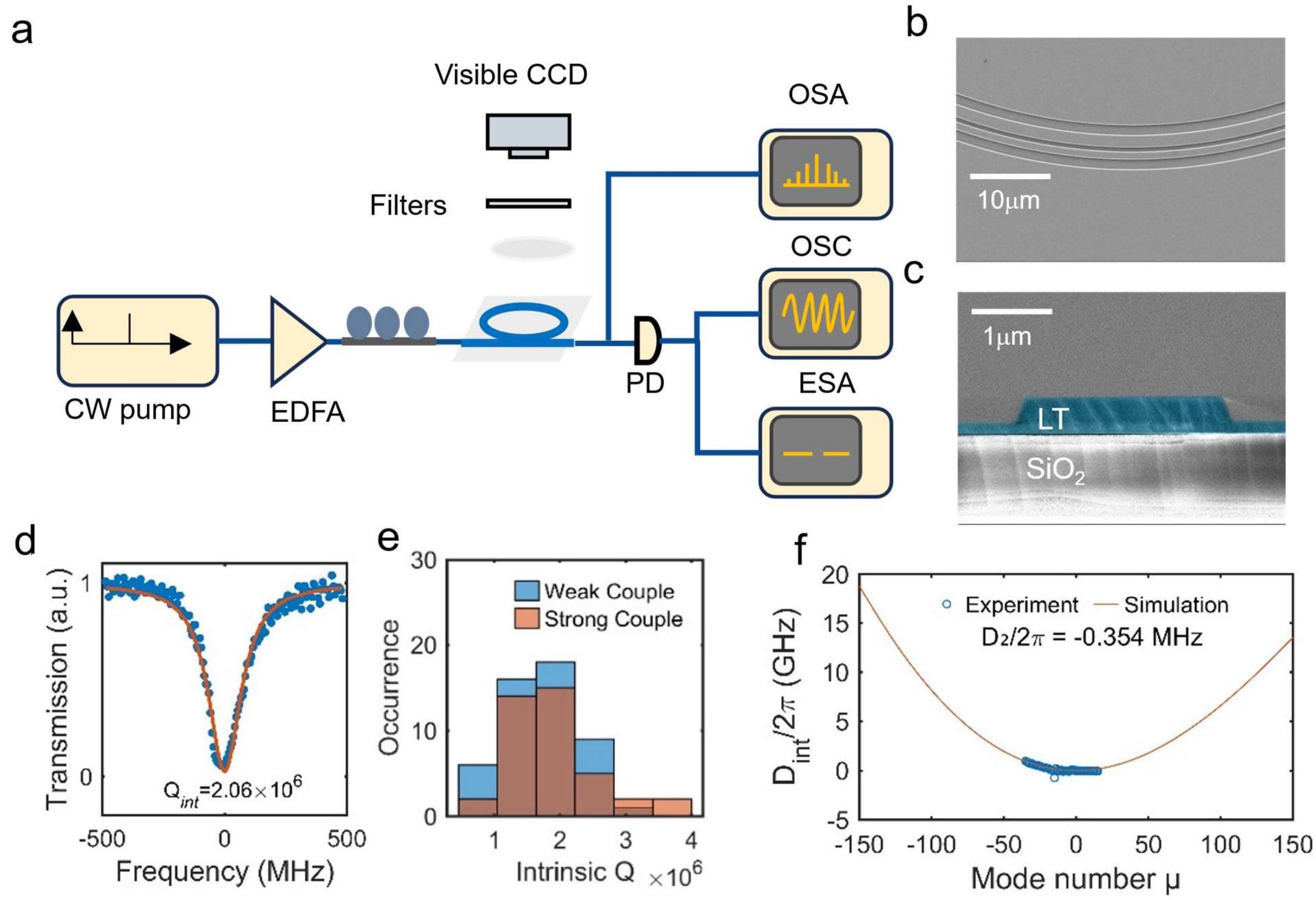

**Fig. 3 | Characterization of a dissipation-engineered TFLT microring resonator. a,** Experimental setup for device characterization. CW, continuous wave; EDFA, erbium-doped fiber amplifier; OSA, optical spectrum analyzer; OSC, oscilloscope; ESA, electrical spectrum analyzer; PD, photodetector. **b,** Scanning electron microscope (SEM) image of the pulley-type coupling region. **c,** SEM image of the waveguide cross-section. **d,** Lorentzian fit to a representative resonance, yielding an intrinsic quality factor $Q_{\mathrm{int}}$=2.06×$10^6$. **e,** Statistical distribution of intrinsic quality factors for weakly coupled and strongly coupled devices. **f,** Measured integrated dispersion $D_{\mathrm{int}}$ and the simulated integrated dispersion of the designed waveguide. The fitted second-order dispersion is $D_2/2\pi$ = -0.354 MHz.

## Generation of microcombs

We demonstrate four distinct frequency comb states by tuning the pump light in four microrings with different parameters, as shown in Fig. 4. These four microrings share the same ring resonator design but differ in their coupling conditions, including the coupling gap, bus waveguide width, and coupling angle. The significant different spectral states observed across the devices highlight the critical role of the dissipation profile, defined by the coupling conditions, in governing the Raman–Kerr dynamics.

In a typical weakly coupled device (device 1, with the structural parameters of Fig. 2a: coupling angle 30°, bus width 1.3 μm, gap 0.86 μm), pumping at 1551.9 nm into resonance generates a frequency comb spanning ~450 nm at an on-chip pump power of ~ 50 mW (Fig. 4a). Further increasing the pump power does not substantially alter the overall comb profile. The transmission curve exhibits a distinct step feature, signaling the transition into a stable Kerr-comb regime (Supplementary Note 3&4). The spectrum is asymmetric. The comb envelope is broader on the long-wavelength side and shows a distinct Raman peak near 1161 nm (Stokes side, corresponding to a Raman shift of ~207 $cm^{-1}$), while the component at a shift of 599 $cm^{-1}$ is strongly suppressed (Detailed Raman analysis is provided in Supplementary Note 2). These spectral characteristics are consistent with a Raman-modified comb envelope, but the overall nonlinear dynamics remain predominantly driven by the Kerr process, and the Raman response does not alter the Kerr comb state. Linewidth measurements of comb lines under different tuning conditions reveal a significant narrowing when transitioning from the modulation-instability (MI) regime to a stable Kerr-comb state. (Supplementary Note 5). Based on the configuration of Fig. 2a, we performed a more detailed gap tuning; in these devices, although the coupling gap varies, all devices consistently produce Kerr combs.

In the device set corresponding to Fig. 2b (coupling angle 45°, bus width 1.25 μm), a relatively weak coupling (gap = 0.95 μm) still yields a Kerr comb with weak Raman perturbation (Fig. 4b). Upon further increasing the coupling (by reducing the gap to 0.83 μm), the device exhibits only Raman lasing, with most of the pump power converted via Raman scattering into Stokes components. Reducing the coupling gap to 0.77 μm yields two representative comb states under different pumping conditions, alongside extra comb lines near the Raman shift branches (Fig. 4d). Pumped at 1560 nm, the device exhibits a pronounced Stokes comb, with both Stokes and anti-Stokes components and a weak Kerr response near the pump. At 1549.8 nm, only the Stokes response and its cascaded orders appear, forming a fully developed dual Raman comb, a phenomenon first observed in an LT microring.

A fully extended Raman–Kerr comb is obtained with a coupling angle of 45°, a bus waveguide width of 1.25 μm and a coupling gap of 0.68 μm (Fig. 4e). The output initially contains the pump and discrete Raman lines: Stokes peaks at ~1710 nm and 1796 nm, and anti-Stokes features at ~1194 nm and 1300 nm. As the pump is tuned further into the red-detuned region, comb lines first appear near the Raman branches,

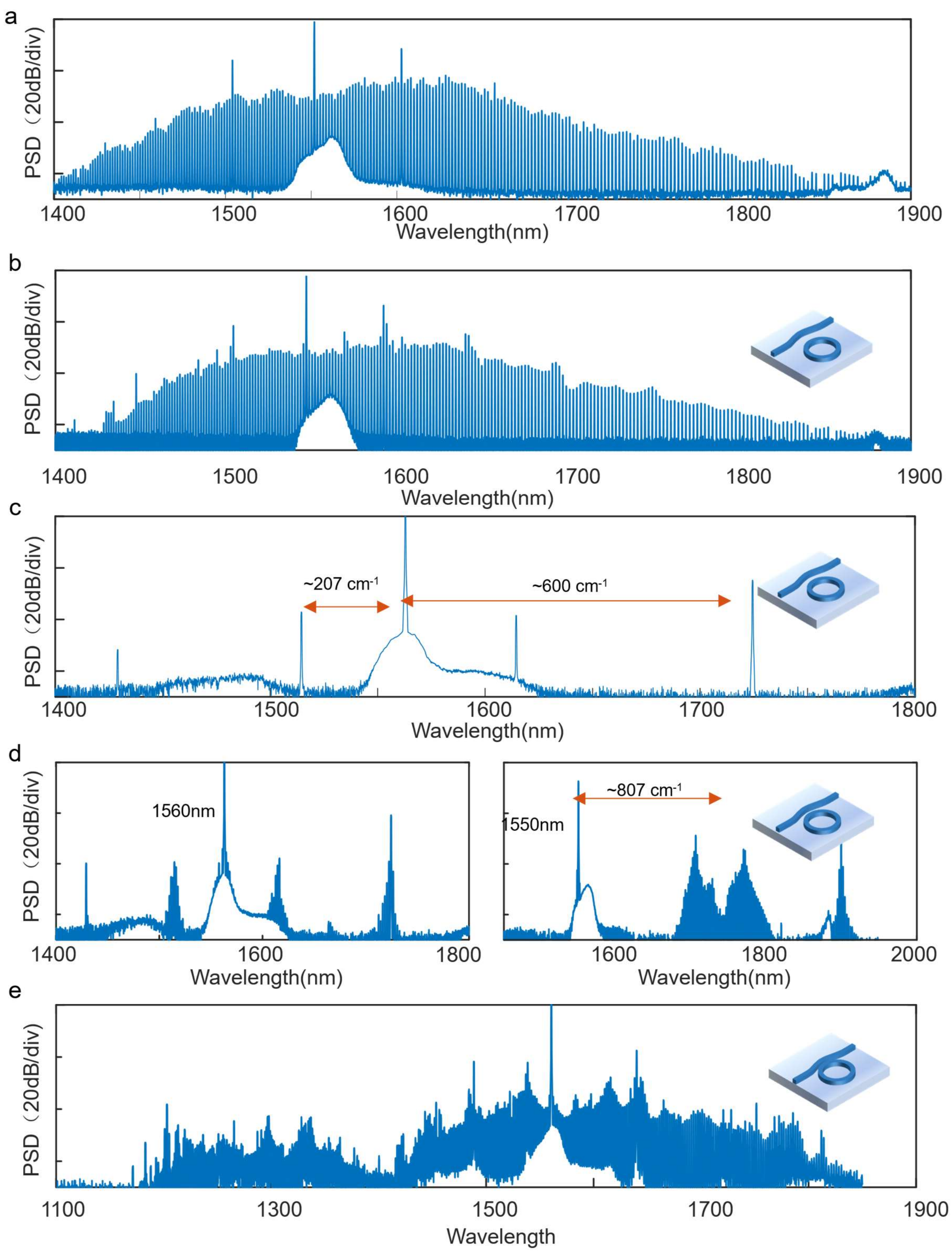

**Fig. 4 | Coupling-controlled frequency comb states in an X-cut TFLT microring. a.** Measured spectra of frequency comb obtained from a weakly coupled device (θ = 30°, gap = 0.86 μm, $w_{bus}$ = 1.3 μm), exhibiting a Kerr comb with weak Raman perturbation. This corresponds to the typical result for the parameter set in the calculations of Fig. 2a. **b-e.** Measured spectra of the microring for parameter (θ = 45°, $w_{bus}$ = 1.25 μm) set in the calculations of Fig. 2b. **b.**

Under a relatively weak coupling condition (gap = 0.95 μm), the device also exhibits a Kerr comb with weak Raman perturbation. **c.** Under a relatively strong coupling condition (gap = 0.83 μm), a Stokes peak appears, indicating excitation of the intrinsic Raman response of the material. **d.** As the coupling is further increased (gap = 0.77 μm), the Raman response becomes stronger, leading to the coexistence of a Raman comb and a Kerr comb (left panel). Alternatively, by changing the pump wavelength, cascaded Raman channels emerge, yielding a pure dual-Raman comb. **e.** In the stronger coupling (gap = 0.68 μm), both the Kerr and Raman thresholds are simultaneously satisfied. Moreover, high-energy higher-order Raman channels open, ultimately generating an ultrabroadband Raman–Kerr comb with a spectral span of approximately 650 nm.

indicating that Raman oscillation seeds local comb formation. With increasing detuning, the Kerr comb centered at the pump gradually develops and fills the spectral gaps between the Raman-dominated branches (Supplementary Note 6). The final state spans about 650 nm(~1190-1840 nm), with pronounced Stokes and anti-Stokes broadening superimposed on a broadband Kerr comb background, resulting in a broadband Raman–Kerr comb.

Together with the coupling analysis in Fig. 2, these experimental results demonstrate that progressively increasing the external coupling drives the comb from a Kerr-comb-dominated regime into a Stokes-assisted hybrid state, and ultimately into a broadband Raman–Kerr synergistic regime. This indicates that different pump conditions combined with variations in the coupling rate enable fine control of the Raman response. Our approach not only yields a wide Kerr comb but also realizes a Raman-assisted ultra-broadband comb. Our work thus extends previous coupling engineering: whereas earlier studies used dissipation to suppress Raman scattering, the stronger loading here retains Raman and transforms it into a cooperative Raman–Kerr state.

## Generation of visible emission

We show multi-wavelength visible emission from a strongly coupled X-cut LTOI microring under continuous-wave pumping at 1556.55 nm as shown in Fig. 5. In contrast, the weakly coupled device in Fig. 4a produces only the third-harmonic light (green, 518.8 nm). The CCD image in Fig. 5a resolves violet (389.1 nm), green (518.8 nm), yellow (~580 nm), orange (603 nm), red (663 nm and 690 nm), and deep red (725 nm) colours. The spectrum in Fig. 5b reveals multiple narrow-linewidth emission peaks in the 450–850 nm range. When the pump wavelength is tuned from 1556.55 nm to 1556.68 nm, these spectral lines are observed reproducibly, confirming that they originate from cavity-enhanced coherent nonlinear frequency conversion rather than thermal radiation or random fluorescence (see Supplementary note 7).

Apart from the pump second harmonic, the spectral lines in the 580–725 nm range cannot be accounted for by simple harmonic generation; instead, they originate from the Raman response of TFLT under strong coupling. Raman measurements reveal dominant phonon modes at ~207 $cm^{-1}$ and ~599 $cm^{-1}$, yielding first-order Stokes wavelengths of 1608 nm and 1717 nm, whose cascaded combination (807 cm$^{-1}$) generates a second-order Stokes component at 1780 nm (Supplementary Note 2). These near-infrared Raman components form a dense spectral reservoir that enables cascaded $\chi^{(2)}$ frequency conversion into the visible. Sum-frequency mixing of the pump (1556 nm) with the 1717 nm Stokes component yields light at ~816 nm, while mixing with anti-Stokes components in the 1200-1300 nm range produces red emission at ~680–710 nm, consistent with the observed 690 nm and 725 nm lines. More notably, the yellow (~580 nm) and orange (~603 nm) emission cannot arise from a single sum-frequency step, but instead require higher-order cascaded conversion. These visible lines originate from cascaded $\chi^{(2)}$ processes involving the second harmonic (778 nm) and higher-order Raman sidebands, or from sum-frequency mixing between the pump and anti-Stokes components under specific phase-matching conditions. Overall, based on spectra and CCD images, eight distinct visible bands are identified; on-chip yellow and orange emission under single-pump excitation is reported here for the first time.

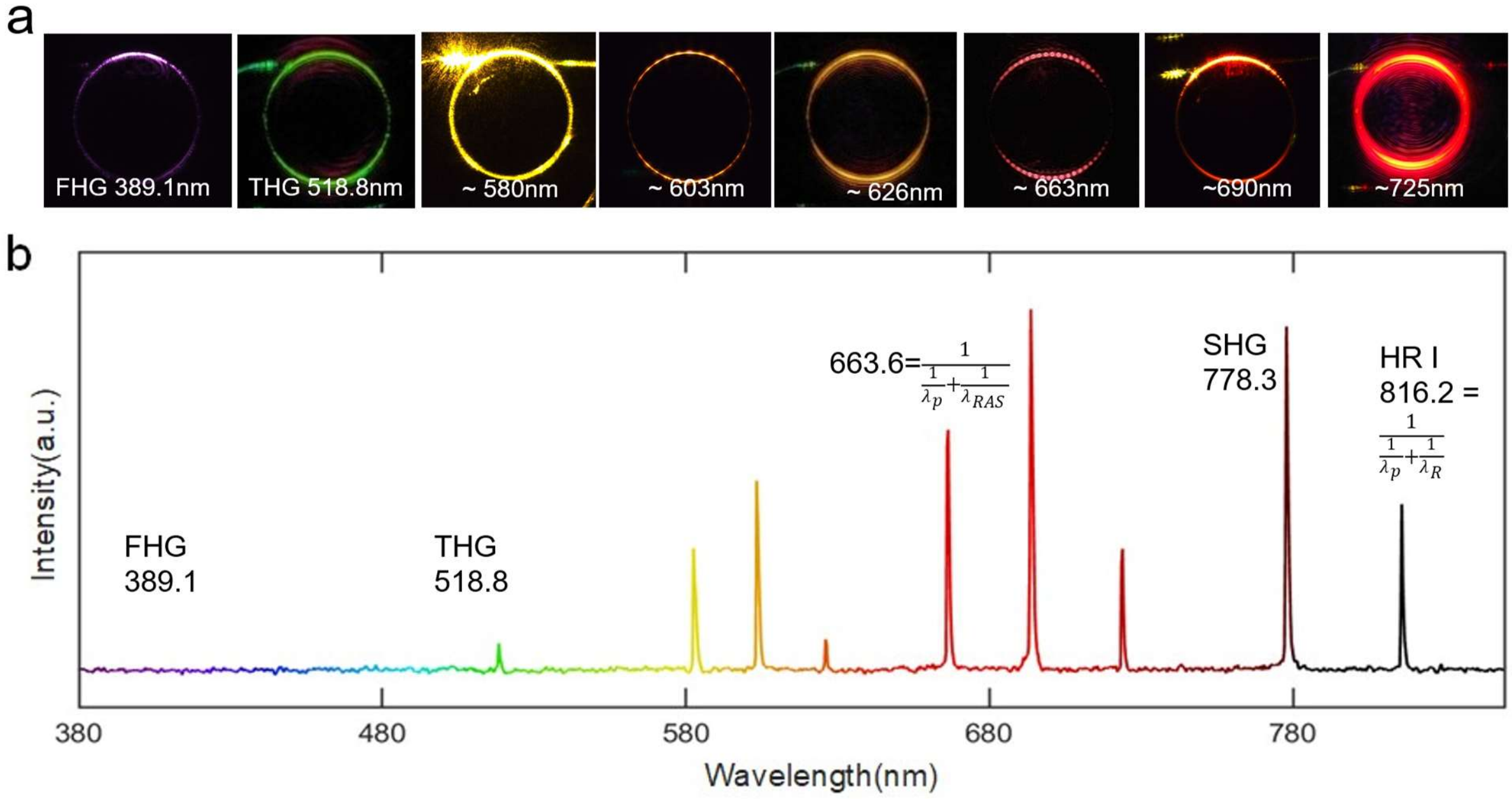

**Fig. 5 | Multi-wavelength emission from a TFLT microring resonator under continuous-wave pumping. a,** CCD image of a microring with strong external coupling, pumped at ≈1556 nm. Using optical bandpass/bandstop filters, we resolve the third harmonic (518.8 nm), the fourth harmonic (389.1 nm), and additional visible emissions at ≈580, 603, 663, 690 and 725 nm. Limited filter bandwidth causes colour mixing in some images (e.g., red light near 690 and 725 nm). **b,** Measured emission spectrum in the visible and near-infrared range (450–850 nm). Combined with the CCD images, eight visible bands are identified: seven bands appear in the spectrum, and the fourth harmonic (389.1 nm) is resolved separately in the images.

These results demonstrate that a strongly coupled X-cut TFLT microring not only generates a near-infrared

Raman–Kerr comb, but also up-converts the energy to multiple visible bands via $\chi^{(2)}$ nonlinearity. The underlying mechanism is that coupling engineering reshapes the loading loss and the intracavity energy distribution among the pump, Raman Stokes and anti-Stokes, and Kerr sideband modes, thereby establishing synergistic Raman–Kerr frequency components. These near-infrared components then undergo cascaded sum-frequency mixing with the pump and its harmonics, yielding polychromatic emission that spans from violet to deep red. Thus, under suitable dispersion and dissipation conditions, a strong Raman response is no longer a parasitic competitor to the Kerr comb, but instead acts as a key bridge linking the near-infrared comb to polychromatic visible generation.

Table 1 | Performance metric comparison with other microcombs on X-cut TFLN/TFLT.

| Platform | Reference | Turn-key potential | Visible emission† | Span(nm) | Optical power consumption(mW) |
|---|---|---|---|---|---|
| X-cut TFLN | Bright Kerr soliton[26] | Yes* | No | 120 | 34 |
| | Raman-Kerr comb[30] | No | No | 300 | 400 |
| | Normal-dispersion Raman-Kerr comb[27] | Yes | No | 330 | ~75 |
| | Normal-dispersion Kerr comb[27] | Yes | No | 210 | ~8 |
| | Bright Kerr soliton[23] | Yes* | No | 200 | ~193 |
| X-cut TFLT | Bright Kerr soliton[28] | Yes | No | 300 | ~90 |
| | Bright Kerr soliton[24] | No | No | 300 | ~110 |
| | Bright Kerr soliton[31] | Yes | No | 240 | ~90 |
| | **Bright Kerr soliton This work** | **Yes‡** | **Yes** | **450** | **~50** |
| | **Raman-Kerr comb This work** | **No** | **Yes** | **650** | **~500** |

*While yet to be demonstrated on X-cut TFLN, previous studies have shown that bright Kerr solitons can exhibit turn-key operation when a DFB laser is self-injection locked to the Kerr micro resonator. ‡For the bright Kerr soliton in this work, a stable comb is reliably obtained by a one-time manual tuning of a CW laser onto resonance (no active feedback thereafter). A DFB laser can also pump the comb, although the generated spectral lines are currently not fully developed, with further optimization, self-injection locking of a DFB laser is promising.† Visible light generation has also been reported on other platforms (e.g., Z-cut LN[32,33], SiC[34],SiN[35], and silica microresonators[36] via distinct

mechanisms: harmonic generation (isolated lines) or pure $\chi^{(3)}$ Kerr combs (without $\chi^{(2)}$). This work demonstrates a Raman–Kerr synergistic visible emission on X-cut LT for the first time.

Comparison of the X-cut TFLT device demonstrated in this work with representative X-cut TFLN and X-cut TFLT microcomb studies is presented in Table 1. Notably, previously reported Kerr combs on X-cut TFLN platforms exhibit spectral spans of only 120–300 nm and show no visible emission. In contrast, the Raman–Kerr frequency comb achieved on the X-cut TFLT platform in this work features a repetition rate of 200 GHz, a spectral span of 650 nm, and simultaneous visible emission, highlighting the advantage of synergistic Raman–Kerr dynamics for broadband polychromatic light generation. Furthermore, the bright Kerr soliton state on the same TFLT device attains a spectral span of 450 nm at an on-chip pump power of only 50 mW, indicating that the TFLT platform can simultaneously support efficient Kerr comb formation and Raman-assisted spectral broadening. Visible light has also been reported on other platforms (e.g., Z-cut TFLN[32], 4H-SiC[34], and silica[36]), but those rely on harmonic generation or pure $\chi^{(3)}$ Kerr combs without $\chi^{(2)}$ assistance, whereas this work demonstrates a broadband visible comb via Raman–Kerr synergy on X-cut TFLT.

## Discussion

In summary, we have demonstrated a coupling-defined dissipation engineering strategy that precisely tailors coupling structures, leaving intrinsic microring dispersion unaltered, to achieve on-demand control over Raman processes. This approach enables selective suppression of Raman scattering for pure Kerr combs, the harnessing of Raman gain for pure Raman combs, or the generation of Raman-assisted ultrawide combs. In X-cut thin-film lithium tantalate microrings, we achieve a Kerr comb spanning 450 nm. By further enhancing the Raman response, we generate a Raman–Kerr synergistic comb spanning 650 nm, representing the broadest combs reported on X-cut TFLN/TFLT platforms to the best of our knowledge. Moreover, in strongly coupled devices, a single near-infrared pump excites visible emission in up to eight bands (from violet to red) via cascaded $\chi^{(2)}$ conversion of the Raman–Kerr comb lines. While the current Raman-Kerr comb exhibits spectral flatness and smoothness that can be further improved, we anticipate that reducing insertion loss and optimizing coupling-region design will enhance spectral uniformity and conversion efficiency. This work fundamentally demonstrates that, through dissipation engineering, a strong Raman response can be transformed from a parasitic limitation into an enabling resource. Looking forward, the multi-band visible emission demonstrated here holds promise for chip-scale atomic clocks and dual-comb spectroscopy, where multiple precisely spaced optical frequencies are essential. This work opens new avenues for comb broadening and polychromatic visible light generation, establishing a new basis for the monolithic integration of nonlinear comb sources, electro-optic modulation, and visible-light generation on X-cut thin-film TFLT platforms.

## Methods

### Numerical simulations

We simulated the intracavity dynamics of the X-cut lithium tantalate microring using a generalized Lugiato–Lefever equation (gLLE) that includes a delayed Raman response. The normalized equation is solved in the rotating frame:

$$\begin{aligned}\frac{\partial \psi}{\partial \tau} &= F^{-1}\{[-\ell(\mu) - i d_{\text{int}}(\mu) + i\zeta]\tilde{\psi}_{\mu}\} \\ &+ i(1-f_R)|\psi|^2\psi + i f_R \psi \int_0^{\infty} h_R(\tau')|\psi(\tau-\tau')|^2 d\tau' + f,\end{aligned} \tag{1.2}$$

where $\psi(\theta, \tau)$ is the field envelope, $F^{-1}$ the inverse Fourier transform, $\mu$ the mode index, $l(\mu)=\kappa(\mu)/\kappa_0$ the normalized total loss, $d_{\text{int}}(\mu)=2D_{\text{int}}(\mu)$ the normalized integrated dispersion, $\zeta$ the normalized detuning, $f_{\text{R}}$ the Raman fraction, and $f$ the normalized pump amplitude.

The measured integrated dispersion $D_{\text{int}}(\mu)$ (from a separate measurement) and the wavelength-dependent external coupling $\kappa_{ex}$ ($\lambda$) obtained from simulations are directly imported. The total loss $\kappa(\mu)= \kappa_i + \kappa_{ex}(\mu)$, combines a constant intrinsic loss and the interpolated coupling loss. The Raman response $h_R(t)$ is a weighted sum of two causal damped oscillators.

The equation is integrated using the split-step Fourier method with 512 modes ($\mu$=-256 to 255). The nonlinear step is advanced with a fourth-order Runge–Kutta scheme and a normalized time step. The pump laser is scanned from low to high detuning following a five-segment linear protocol. Further details are provided in Supplementary Note 1.

### Device fabrication

The devices were fabricated on a lithium tantalate on insulator (LTOI) wafer consisting of a silicon substrate, a 9-μm-thick buried $SiO_2$ layer, and a 600-nm-thick X-cut single-crystal thin film lithium tantalate (TFLT) thin film. A 600-nm-thick amorphous silicon (a-Si) layer was first deposited on the wafer surface by plasma-enhanced chemical vapor deposition (PECVD) to serve as a hard mask. Subsequently, AR-P 6200.09 electron-beam resist was spin-coated onto the sample. Microring resonator patterns were defined in the resist using electron-beam lithography (Vistec EBPG-5200+), followed by pattern transfer into the a-Si hard mask via inductively coupled plasma (ICP) etching. Using the patterned a-Si layer as an etch mask, the underlying TFLT film was etched to a depth of 450 nm, forming the microring waveguide structures. After etching, the sample was immersed in a KOH: water (1:1) solution at 80 °C to remove the residual a-Si hard mask, resulting in the completed on-chip LT devices.

**Author contributions**

†These authors contributed equally: Xin Wang, Mingkun Xiao, and Min Sun. X.W. and M.X. conceived the idea. M.S. and X.W. performed the device fabrication. Transmission spectrum measurements, pump experiments, and soliton-step characterization were carried out by X.W. and M.X. Simulations were conducted by X.W. with assistance from M.X. Visible-light characterization was performed by R.G. and X.W. Linewidth measurements were performed by Y.C. and X.W. Data processing was performed by X.W. and M.X. X.J. and Y.Z. provided valuable guidance on the experiments and manuscript preparation. J.L., Y.S., X.J., and Y.Z. supervised the project. All authors contributed to writing the manuscript.

**Acknowledgements**

This work was supported by the National Natural Science Foundation of China (NSFC) under Grant 62335014 and 62575169, the Shanghai Municipal Science and Technology Commission Project under Grant 25JD1402000 and 25ZR1401178. We acknowledge the Center for Advanced Electronic Materials and Devices (AEMD) at Shanghai Jiao Tong University (SJTU).

**Competing interests**

The authors declare no competing interests.

**Data availability**

The data supporting the plots in this study and other findings of this study are available from the corresponding author upon reasonable request.